\documentclass[11pt,twoside]{article}
\usepackage[pdftex]{graphicx}
\usepackage{amsmath}
\usepackage{amssymb}
\usepackage{cite}

\begin{document}
\newcommand{\pst}{\hspace*{1.5em}}

\newcommand{\rigmark}{\em Journal of Russian Laser Research}
\newcommand{\lemark}{\em Volume 30, Number 5, 2009}

\newcommand{\be}{\begin{equation}}
\newcommand{\ee}{\end{equation}}
\newcommand{\bm}{\boldmath}
\newcommand{\ds}{\displaystyle}
\newcommand{\bea}{\begin{eqnarray}}
\newcommand{\eea}{\end{eqnarray}}
\newcommand{\ba}{\begin{array}}
\newcommand{\ea}{\end{array}}
\newcommand{\arcsinh}{\mathop{\rm arcsinh}\nolimits}
\newcommand{\arctanh}{\mathop{\rm arctanh}\nolimits}
\newcommand{\bc}{\begin{center}}
\newcommand{\ec}{\end{center}}

\thispagestyle{plain}

\label{sh}


\begin{center} {\Large \bf
\begin{tabular}{c}
CONDITIONAL INFORMATION\\[-1mm]
AND HIDDEN CORRELATIONS\\[-1mm]
IN SINGLE-QUDIT STATES
\end{tabular}
 } \end{center}

\bigskip

\bigskip

\begin{center} {\bf
Margarita A. Man'ko }\end{center}

\medskip

\begin{center}
{\it
Lebedev Physical Institute, Russian Academy of Sciences\\
Leninskii Prospect 53, Moscow 119991, Russia}

\smallskip

$^*$E-mail:~~~mmanko\,@\,sci.lebedev.ru\\
\end{center}

\begin{abstract}\noindent
We discuss the notions of mutual information and conditional
information for noncomposite systems, classical and quantum; both
the mutual information and the conditional information are
associated with the presence of hidden correlations in the state of
a single qudit. We consider analogs of the entanglement phenomena in
the systems without subsystems related to strong hidden quantum
correlations.
\end{abstract}

\medskip


\section{Introduction}
\pst
The main goal of this work is to show that systems without
subsystems and multipartite systems, both classical and quantum,
have identical correlation properties; the difference between the
systems is related to interpretation of the correlations. We call
the correlations in systems without subsystems the hidden
correlations.

The probability distribution $P(k)\geq 0$ $(k=1,2,\ldots,N)$ for one
random variable is the normalized function $\sum_{k=1}^NP(k)=1$,
which characterizes the state of a classical finite
system~\cite{Kolmogorov-book,Holevo-book,Chuang-book}. If the
classical system contains two subsystems, the joint probability
distribution ${\cal P}(j,k)\geq 0$, where $j=1,2,\ldots,N_1$,
$k=1,2,\ldots,N_2$, and $N=N_1N_2$, characterizes the statistical
properties of the system with two random variables. The
normalization condition for the joint probability distribution of
two random variables $\sum_{j=1}^{N_1}\sum_{k=1}^{N_2}{\cal
P}(j,k)=1$ is the condition for $N$ nonnegative numbers ${\cal
P}(j,k)$ organized in the form of a table analogous to a rectangular
matrix with the $j$th row and the $k$th column.

In the case of a finite tripartite classical system, the statistical
properties of the system with three random variables are described
by a joint probability distribution $\Pi(j,k,l)$, where
$j=1,2,\ldots,N_1$, $k=1,2,\ldots,N_2$, and $\ell=1,2,\ldots,N_3$,
and the probability distribution satisfies the normalization
condition $\sum_{j=1}^{N_1}\sum_{k=1}^{N_2}\sum_{\ell=1}^{N_3}
\Pi(j,k,\ell)=1$; we assume that $N_1N_2N_3=N$.

For quantum systems, the states are described by the density
matrices~\cite{Landau,vonNeumann,vonNeumann-book} or by the
tomographic probability distributions (see, for
example,~\cite{Ibort-PS150}).

For the single-qudit state or the spin state, the density matrix
$\rho_{mm'}$, where $m,m'=-j,-j+1,\ldots,j-1,j$, is the Hermitian
nonnegative matrix $\rho=\rho^\dagger$ with unit trace Tr$\,\rho=1$
and nonnegative eigenvalues. For a bipartite system containing two
qudits, the density matrix $\rho_{m_1m_2,\,m'_1m'_2}$, where
$m_1,m'_1=-j_1,-j_1+1,\ldots,j_1-1,j_1$ and
$m_2,m'_2=-j_2,-j_2+1,\ldots,j_2-1,j_2$, has nonnegative
eigenvalues, the property of hermiticity, and unit trace.
Analogously, for the system of three qudits, the density matrix of
the system state $\rho_{m_1m_2m_3,\,m'_1m'_2m'_3}$, where $m_a,m'_a=
-j_a,-j_a+1,\ldots,j_a-1,j_a$ $(a=1,2,3)$, is the nonnegative
Hermitian matrix with unit trace.

If one makes an appropriate map of indices in the density matrices,
the matrix elements in these matrices can be labeled by integers as
follows: $\rho_{jj'}$~(one qudit), $\rho_{jk,\,j'k'}$~(two qudits),
and $\rho_{jk\ell,\,j'k'\ell'}$~(three qudits), where
$j,j'=1,2,\ldots,N_1$, $k,k'=1,2,\ldots,N_2$, and
$\ell,\ell'=1,2,\ldots,N_3$, The diagonal elements of the density
matrices provide the probability distributions $\rho_{jj}=P(j)$,
$\rho_{jk,\,jk}={\cal P}(j,k)$, and
$\rho_{jk\ell,\,jk\ell}=\Pi(j,k,\ell)$.

As we have discussed in
\cite{1,2,3,NuovoCimC,5,6,7,8,9,10,11,12,13,15,Entropy,21,22},
one can consider probability distributions of composite and
noncomposite systems as a set of $N$ nonnegative numbers; in the
case of $N=N_1N_2$ or $N=N_1N_2N_3$, where $N_1$, $N_2$, and $N_3$
are integers, one can obtain new entropy--information relations for
noncomposite systems. Analogously, for quantum indivisible systems,
one can obtain the relations for entropy and information analogous
to the relations known for multiqudit systems.

The aim of this study is to extend the notions of mutual information
and conditional information available for composite systems to the
single-qudit state in order to employ them as characteristics of
hidden correlations in the system, including the correlations
associated with the entanglement phenomena in single-qudit states.
It is worth noting that strong quantum correlations in single-qudit
states have been studied in \cite{Shumov}. Also we apply the
introduced notion to quantum thermodynamics of $N$-level
atoms~\cite{Castanos}. Recently, new aspects of quantum
thermodynamics were discussed, for example,
in~\cite{Huber,Oppenheim,Julio} and in the talk of
Facchi~\cite{Facchi} presented at the International Workshop on
Foundations of Quantum Mechanics and Applications (Madrid,
30~January\,--\,10~February, 2017).

\section{Correlations and Entropies}
\pst
In this section, we review known properties of correlations in
bipartite and tripartite classical systems and their relations with
the mutual information and conditional information, respectively.

For two classical random variables, the joint probability
distribution ${\cal P}(j,k)$, where $j=1,2,\ldots,N_1$ and
$k=1,2,\ldots,N_2$, determines the marginal probability
distributions
\begin{equation}\label{1}
{\cal P}_1(j)=\sum_{k=1}^{N_2} {\cal P}(j,k),\qquad {\cal
P}_2(k)=\sum_{j=1}^{N_1} {\cal P}(j,k)
\end{equation}
describing the statistical properties of the first and second
subsystems, respectively.

By definition, we have three Shannon entropies associated with three
probability distributions ${\cal P}(j,k)$, ${\cal P}_1(j)$ and
${\cal P}_2(k)$; they read
\begin{eqnarray}\label{2}
H(1,2)=-\sum_{j=1}^{N_1}\sum_{k=1}^{N_2}{\cal P}(j,k)\ln {\cal
P}(j,k),\quad H(1)=-\sum_{j=1}^{N_1} {\cal P}_1(j)\ln  {\cal
P}_1(j),\quad H(2)=-\sum_{k=1}^{N_2} {\cal P}_2(k)\ln  {\cal
P}_2(k).\nonumber\\[-2mm]
\end{eqnarray}
If there is no correlations between the degrees of freedom of two
subsystems, the joint probability distribution ${\cal P}(j,k)$ has
the factorized form in terms of the marginal probability
distributions ${\cal P}_1(j)$ and ${\cal P}_2(k)$,
\begin{equation}\label{3}
{\cal P}(j,k)={\cal P}_1(j){\cal P}_2(k).
\end{equation}
Such expression means that the entropy of the system $H(1,2)$ is the
sum of entropies of two subsystems, i.e., the following equality
holds:
\begin{equation}\label{4}
H(1,2)=H(1)+H(2).
\end{equation}

If there are correlations between the degrees of freedom of two
subsystems, one has the inequality for the entropies
\begin{equation}\label{5}
0\leq I=H(1)+H(2)-H(1,2),
\end{equation}
which is the nonnegativity condition for the mutual information $I$.
Thus, the value of mutual information is a characteristic of the
correlations in the system consisting of two subsystems or the
system with two random variables.

Analogously, for tripartite classical system, the joint probability
distribution $\Pi(j,k,\ell)$, $j=1,2,\ldots,N_1$,
$k=1,2,\ldots,N_2$, and $\ell=k=1,2,\ldots,N_3$, describing the
statistical properties and correlations in the system with three
random variables determines the marginal probability distributions
\begin{equation}\label{6}
{\cal P}_{12}^{(\Pi)}(j,k)=\sum_{\ell=1}^{N_3}\Pi(j,k,\ell),\quad
{\cal P}_{23}^{(\Pi)}(k,\ell)=\sum_{j=1}^{N_1}\Pi(j,k,\ell),\quad
P_2^{(\Pi)}(k)=\sum_{j=1}^{N_1}\sum_{\ell=1}^{N_3}\Pi(j,k,\ell).
\end{equation}
These marginal probability distributions for three different
subsystems of tripartite system are characterized by three Shannon
entropies
\begin{eqnarray}\label{7}
&& H^{(\Pi)}(1,2)=-\sum_{j=1}^{N_1}\sum_{k=1}^{N_2}{\cal
P}_{12}^{(\Pi)}(j,k)\ln {\cal P}_{12}^{(\Pi)}(j,k),\qquad
H^{(\Pi)}(2,3)=-\sum_{k=1}^{N_2}\sum_{\ell=1}^{N_3}{\cal
P}_{23}^{(\Pi)}(k,\ell)\ln {\cal
P}_{23}^{(\Pi)}(k,\ell),\nonumber\\[-2mm]
&&\\[-2mm]
&&\qquad\qquad\qquad\qquad\qquad\qquad
H^{(\Pi)}(2)=-\sum_{k=1}^{N_2}P_{2}^{(\Pi)}(k)\ln P_{2}^{(\Pi)}(k).
\nonumber
\end{eqnarray}
If there is no correlations between the degrees of freedom in the
system with three random variables, the joint probability
distribution $\Pi(j,k,\ell)$ has the factorized form in terms of
marginal probability distributions, i.e.,
\begin{equation}\label{8}
\Pi(j,k,\ell)=\left(\sum_{k'=1}^{N_2}{\cal
P}_{12}^{(\Pi)}(j,k')\right)P_2^{(\Pi)}(k)\left(\sum_{k''=1}^{N_2}{\cal
P}_{2,3}^{(\Pi)}(k'',\ell)\right).
\end{equation}
In the case of tripartite classical system with three random
variables, for the conditional information $I_C$ we have the
nonnegativity condition
\begin{equation}\label{9}
0\leq I_C=H^{(\Pi)}(1,2)+H^{(\Pi)}(2,3)-H^{(\Pi)}(2)-H(1,2,3),
\end{equation}
where the entropy of the tripartite-system state reads
\begin{equation}\label{10}
H(1,2,3)=-\sum_{j=1}^{N_1}\sum_{k=1}^{N_2}\sum_{\ell=1}^{N_3}\Pi(j,k,\ell)
\ln \Pi(j,k,\ell).
\end{equation}
For the systems without correlations determined by the joint
probability distribution~(\ref{8}), the conditional information is
equal to zero.

Thus, the value of conditional information $I_C$ is a
characteristics of the correlations in the tripartite system. The
properties of the set of nonnegative numbers ${\cal P}(j,k)$ and
$\Pi(j,k,\ell)$, from the mathematical point of view, are
characterized by the values of numbers $I$ and $I_C$, which show the
difference of two possibilities. The first one is either to
represent the number ${\cal P}(j,k)$ in the product form~(\ref{3})
or to represent the numbers $\Pi(j,k,\ell)$ in the product
form~(\ref{8}) or as $\Pi(j,k,\ell)=P_1(j)P_2(k)P_3(\ell)$.
Information $I$ and conditional information $I_C$ show how much the
sets of numbers ${\cal P}(j,k)$ and $\Pi(j,k,\ell)$ differ from
products of the corresponding marginal probability distributions. In
this formulation, we do not interpret the sets of numbers as
probability distributions but simply consider the sets as tables of
given nonnegative numbers ${\cal P}(j,k)$ and $\Pi(j,k,\ell)$
satisfying the normalization conditions. Thus, the numbers $I$ and
$I_C$ can be associated not only with the joint probability
distributions of bipartite and tripartite systems but also with
abstract sets of nonnegative numbers.

\section{Quantum States of Bipartite and Tripartite Systems}
\pst
For quantum states of any system, one has the density matrix $\rho$
which determines the von Neumann entropy,
\begin{equation}\label{21}
S=-\mbox{Tr}\,\rho\ln\rho.
\end{equation}
For bipartite system with two qudits, the density matrix $\rho(1,2)$
determines the von Neumann entropy,
\begin{equation}\label{22}
S(1,2)=-\mbox{Tr}\,\rho(1,2)\ln\rho(1,2).
\end{equation}
The density matrices of the subsystem states of the first and second
qudits are given as
\begin{equation}\label{23}
\rho(1)=\mbox{Tr}_2\rho(1,2),\qquad \rho(2)=\mbox{Tr}_1\rho(1,2),
\end{equation}
and the von Neumann entropies of these states
\begin{equation}\label{24}
S(1)=-\mbox{Tr}\,\rho(1)\ln\rho(1),\qquad
S(2)=-\mbox{Tr}\,\rho(2)\ln\rho(2)
\end{equation}
satisfy the nonnegativity condition. The mutual quantum information
$I_q$ is given by the relationship
\begin{equation}\label{25}
0\leq I_q=S(1)+S(2)-S(1,2).
\end{equation}
For the system without quantum correlations between the subsystem
degrees of freedom, one has $I_q=0$ and, in this case, the density
matrix has the factorized form
\begin{equation}\label{26}
\rho(1,2)=\rho(1)\otimes\rho(2).
\end{equation}
We consider the system of three qudits with the density matrix
$\rho(1,2,3)$ and three density matrices of subsystems $\rho(1,2)$,
$\rho(2,3)$, and $\rho(2)$, respectively; the three density matrices
are obtained from the density matrix $\rho(1,2,3)$ using partial
traces, namely,
\begin{equation}\label{27}
\rho(1,2)=\mbox{Tr}_3\,\rho(1,2,3),\qquad\rho(2,3)=\mbox{Tr}_1\,\rho(1,2,3),\qquad
\rho(2)=\mbox{Tr}_1\,\rho(1,2).
\end{equation}
The conditional quantum information $I_{Cq}$ is defined in terms of
von Neumann entropies of the subsystem states; it reads
\begin{equation}\label{28}
I_{Cq}=S(1,2)+S(2,3)-S(1,2,3)-S(2)
\end{equation}
and satisfies the nonnegativity condition $I_{Cq}\geq 0.$ For the
system states without correlations such that
$\rho(1,2,3)=\rho(1)\otimes\rho(2)\otimes\rho(3)$, the conditional
quantum information $I_{Cq}=0$. Thus, the value of conditional
quantum information characterizes the degree of quantum correlations
in the tripartite system.

Now we express the von Neumann entropies~(\ref{22}) and (\ref{23})
in terms of the matrix elements of the density matrices
$~\rho_{jk,\,j'k'}(1,2)=\langle jk\mid\widehat\rho(1,2)\mid
j'k'\rangle$, $~\rho_{jj'}(1)=\langle j\mid\widehat\rho(1)\mid
j'\rangle$, and $~\rho_{kk'}(2)=\langle k\mid\widehat\rho(2)\mid
k'\rangle$ of the corresponding density operators
$\widehat\rho(1,2)$, $\widehat\rho(1)$, and $\widehat\rho(2)$ of the
qudit states. We have
\begin{equation}\label{29}
\rho_{jj'}(1)=\sum_{k=1}^{N_2}\rho_{jk,\,j'k}(1,2),\qquad
\rho_{kk'}(2)=\sum_{j=1}^{N_1}\rho_{jk,\,jk'}(1,2).
\end{equation}
The von Neumann entropies~(\ref{24}) of the qudit states read
\begin{equation}\label{30}
S(1)=-\sum_{j=1}^{N_1}\big[\rho(1)\ln\rho(1)\big]_{jj}\,,\qquad
S(2)=-\sum_{k=1}^{N_2}\big[\rho(2)\ln\rho(2)\big]_{kk}\,,
\end{equation}
and the von Neumann entropy of the bipartite system states is
\begin{equation}\label{31}
S(1,2)=-\sum_{j=1}^{N_1}\sum_{k=1}^{N_2}
\big[\rho(1,2)\ln\rho(1,2)\big]_{jk,\,jk}\,.
\end{equation}
Analogous expressions can be easily obtained for the von Neumann
entropies of tripartite system states.

The idea of our approach to introduce the notion of conditional
information and the notion of mutual information for single qudit
states is related to the employment of bijective maps of indices
determining the matrix elements of density matrices. We describe the
map in the next section.

\section{The Functions Detecting the Hidden Correlations}
\pst
We introduce the sets of functions describing the bijective map of
integers $y=1,2,\ldots,N$, where $N=X_1X_2$, onto the pairs of
integers $y\leftrightarrow x_1,x_2$, where $x_1=1,2,\ldots,X_1$ and
$x_2=1,2,\ldots,X_2$. Following~\cite{Zhenat-JRLR1,Vova-TMF}, we
obtain
\begin{eqnarray}
&&y(x_1,x_2)=x_1+(x_2-1)X_1,\qquad 1\leq x_1 \leq X_1,\qquad 1\leq
x_2\leq X_2,\nonumber\\[-2mm]
\label{A}\\[-2mm]
&&x_1(y)=y\,\mbox{mod}\,{X_1},\qquad
x_2(y)-1=\frac{y-x_1(y)}{X_1}\,\mbox{mod}\,{X_2},\qquad 1\leq y
\leq N.\nonumber
\end{eqnarray}
It is worth noting that analogous functions were discussed in
\cite{Antonella-arXiv,FacchiJPA}. In the case where the integer $N$ is the
product of three integers $N=X_1X_2X_3$, we introduce the functions
$y(x_1,x_2,x_3)$, $x_1(y)$, $x_2(y)$, and $x_3(y)$ given by the
expressions~\cite{Zhenat-JRLR1}
\begin{eqnarray}
&&y=y(x_1,x_2,x_3)=x_1+(x_2-1)X_1+(x_3-1)X_1 X_2,\qquad 1\leq x_i
\leq X_i,\qquad i\in[1,3],\nonumber\\
&&x_1(y)=y\,\mbox{mod}\,{X_1},\qquad
x_2(y)-1=\frac{y-x_1(y)}{X_1}\,\mbox{mod}\,{X_2},  \label{B}\\
&& x_3(y)-1=\frac{y-x_1(y)-(x_2(y)-1)X_1}{X_1
X_2}\,\mbox{mod}\,{X_3}.\nonumber
\end{eqnarray}
The introduced functions provide the bijective map of integers onto
pairs of integers and triples of integers. These maps give the
possibilities to interpret the probability distributions of one
random variable $P(y)$, $y=1,2,\ldots,N$ as the probability
distributions of two random variables ${\cal P}(x_1,x_2)\equiv
P\big(y(x_1,x_2)\big)$ if $N=N_1N_2.$

If $N=N_1N_2N_3$, the functions introduced provide the possibility
to interpret the probability distributions of one random variable
$P(y)$ as joint probability distributions $\Pi(x_1,x_2,x_3)\equiv
P\big(y(x_1,x_2,x_3)\big)$ of three random variables. In view of
such interpretation, we introduce the notion of mutual information
and the notion of conditional information for single qudit states.

First, we introduce the notion of artificial subsystems in the case
of classical system described by the probability distribution
$P(y)$, $y=1,2,\ldots,N$ and $N=N_1N_2$. We construct two marginal
probability distributions ${\cal P}_1(x_1)$ and ${\cal P}_2(x_2)$,
\begin{equation}\label{42}
{\cal P}_1(x_1)=\sum_{x_2=1}^{N_2}P\big(y(x_1,x_2)\big),\qquad {\cal
P}_2(x_2)=\sum_{x_1=1}^{N_1}P\big(y(x_1,x_2)\big).
\end{equation}
The von Neumann entropies $H_a(1)$ and $H_a(2)$ for artificial
subsystem states read
\begin{equation}\label{43}
H_a(1)=-\sum_{x_1=1}^{N_1}{\cal P}_1(x_1)\ln {\cal P}_1(x_1),\quad
H_a(2)=-\sum_{x_2=1}^{N_2}{\cal P}_2(x_2)\ln {\cal P}_2(x_2).
\end{equation}
The mutual information $I_a$ for the classical system of one random
variable and the two artificial subsystems introduced is
\begin{equation}\label{44}
I_a=H_a(1)+H_a(2)+\sum_{x_1=1}^{N_1}\sum_{x_2=1}^{N_2}
P\big(y(x_1,x_2)\big)\ln P\big(y(x_1,x_2)\big)\geq 0.
\end{equation}
The value of information $I_a$ characterizes the degree of hidden
correlations in the system with one random variable, and the hidden
correlations demonstrate the difference of the probability
distribution $P\big(y(x_1,x_2)\big)$ and the product of marginals:
$\Delta(x_1,x_2)={\cal P}_1(x_1){\cal
P}_2(x_2)-P\big(y(x_1,x_2)\big)$. In the absence of hidden
correlations, this difference is equal to zero, and the mutual
information $I_a=0$ also.

For the classical system with one random variable and $N=N_1N_2N_3$,
we introduce an analogous construction of three artificial
subsystems and define the marginal probability distributions,
\begin{eqnarray}\label{45}
&&{\cal
P}_{12}^{(a)}(x_1,x_2)=\sum_{x_3=1}^{N_3}P\big(y(x_1,x_2,x_3)\big),
\qquad
{\cal
P}_{23}^{(a)}(x_2,x_3)=\sum_{x_1=1}^{N_1}P\big(y(x_1,x_2,x_3)\big),\nonumber\\[-2mm]
\\[-2mm]
&&\qquad\qquad\qquad\qquad\qquad {\cal
P}_{2}^{(a)}(x_2)=\sum_{x_1=1}^{N_1}\sum_{x_3=1}^{N_3}
P\big(y(x_1,x_2,x_3)\big).\nonumber
\end{eqnarray}
The above probability distributions determine the artificial
subsystem entropies which, in turn, determine the nonnegative
conditional information; it reads
\begin{eqnarray}\label{46}
0\leq I_{ac}=-\sum_{x_1=1}^{N_1}\sum_{x_2=1}^{N_2} {\cal
P}_{12}^{(a)}(x_1,x_2)\ln{\cal P}_{12}^{(a)}(x_1,x_2)
-\sum_{x_1=1}^{N_2}\sum_{x_3=1}^{N_3}{\cal P}_{23}^{(a)}(x_2,x_3)
\ln{\cal P}_{23}^{(a)}(x_2,x_3)\nonumber\\
+\sum_{x_2=1}^{N_2}{\cal P}_{2}^{(a)}(x_2)\ln{\cal P}_{2}^{(a)}(x_2)
+\sum_{x_1=1}^{N_1}\sum_{x_2=1}^{N_2}\sum_{x_1=3}^{N_3}P\big(y(x_1,x_2,x_3)\big)\ln
P\big(y(x_1,x_2,x_3)\big).
\end{eqnarray}
The conditional information is equal to zero if there is no hidden
correlations in the system.

Analogously, we can introduce artificial subsystems for an arbitrary
classical system with one random variable for the case of
$N=N_1N_2\cdots N_n$. For this, we use an appropriate partition of
the integer $N$.

For the quantum system, i.e., a single qudit with the density matrix
$\rho_{yy'}$, where $y,y'=1,2,\ldots N$, we introduce the notion of
mutual quantum information (for $N=N_1N_2$) and the notion of
conditional quantum information (for $N=N_1N_2N_3$) applying an
analogous partition tool. This means that we can interpret the
density matrix $\rho_{yy'}$ as the density matrix of the bipartite
system determining it, namely,
$\rho_{x_1x_2,\,x'_1x'_2}\equiv\rho_{y(x_1x_2),\,y'(x'_1x'_2)}.$ In
view of this definition, we introduce the density matrix of the
first artificial subsystem (the first qudit) as
$\rho^{(a)}_{x_1x'_1}(1)=\sum_{x_2=1}^{N_2}\rho_{y(x_1x_2),\,y'(x'_1x_2)}$,
and the density matrix of the second artificial subsystem (the
second qudit) as
$\rho^{(a)}_{x_2x'_2}(2)=\sum_{x_1=1}^{N_1}\rho_{y(x_1x_2),\,y'(x_1x'_2)}.$
The construction of the density matrices provides the possibility to
introduce the notion of nonnegative mutual quantum information
defined through von Neumann entropies, i.e.,
\begin{eqnarray}\label{47}
&&0\leq I_q^{(a)}=
-\sum_{x_1=1}^{N_1}\big(\rho^{(a)}(1)\ln\rho^{(a)}(1)\big)_{x_1x_1}-
\sum_{x_2=1}^{N_2}\big(\rho^{(a)}(2)\ln\rho^{(a)}(2)\big)_{x_2x_2}\nonumber\\[-2mm]
&&\qquad\qquad\qquad\qquad+\sum_{x_1=1}^{N_1}\sum_{x_2=1}^{N_2}
\big(\rho\ln\rho\big)_{y(x_1,\,x_2),\,y'(x_1,\,x_2)}.
\end{eqnarray}
The value of mutual information characterizes the difference of the
density matrix $\rho_{x_1x_2,\,x'_1x'_2}(1,2)$ and the product of
two density matrices $\rho_{x_1x_2}^{(a)}(1)$ and
$\rho_{x'_1x'_2}^{(a)}(2)$, namely,
\begin{equation}\label{48}
\big(\Delta\rho\big)_{x_1x_2,\,x'_1x'_2}=\rho_{x_1x_2,\,x'_1x'_2}(1,2)-
\big(\rho^{(a)}(1)\otimes \rho^{(a)}(2)\big)_{x_1x_2,\,x'_1x'_2}.
\end{equation}
If there is no hidden correlations in the system of two artificial
qudits, this difference is equal to zero and the mutual quantum
information $I_{q}^{(a)}=0$ also.

For a single-qudit state with spin $s=(N-1)/2$, where the integer
$N=N_1N_2N_3$, the density matrix $\rho_{yy'}$ with indices
$y,y'=1,2,\ldots,N$ can be interpreted as the density matrix of the
tripartite system with three artificial qudits using the definition
$\rho_{x_1x_2x_3,\,x'_1x'_2x'_3}\equiv\rho_{y(x_1,x_2,x_3),\,y'(x'_1,x'_2,x'_3)}$,
where indices $x_1,\,x_2,\,x_3,\,x'_1,\,x'_2,\,x'_3$ take the values
$x_1,x'_1=1,2,\ldots,N_1$, $x_2,x'_2=1,2,\ldots,N_2$, and
$x_3,x'_3=1,2,\ldots,N_3$. Such interpretation provides the
possibility to introduce the density matrices of three artificial
subsystems with density matrices of their states obtained by the
partial traces,
\begin{eqnarray}\label{49}
&&\rho^{(a)}_{x_1x_2,\,x'_1x'_2}(1,2)=\sum_{x_3=1}^{N_3}
\rho_{x_1x_2x_3,\,x'_1x'_2x_3},\qquad
\rho^{(a)}_{x_2x_3,\,x'_2x'_3}(2,3)=\sum_{x_1=1}^{N_1}
\rho_{x_1x_2x_3,\,x_1x'_2x'_3},\nonumber\\[-2mm]
&&\qquad\qquad\qquad\qquad\qquad
\rho^{(a)}_{x_2x'_2}(2)=\sum_{x_1=1}^{N_1}\rho_{x_1x_2,\,x_1x'_2}(1,2).
\end{eqnarray}
After obtaining these matrices, we introduce the conditional quantum
information for the single qudit state using the definition
\begin{eqnarray}\label{50}
0\leq
I_{Cq}^{(a)}&=&-\mbox{Tr}\left\{\rho^{(a)}(1,2)\ln\rho^{(a)}(1,2)\right\}
-\mbox{Tr}\left\{\rho^{(a)}(2,3)\ln\rho^{(a)}(2,3)\right\}\nonumber\\
&&+\mbox{Tr}\left\{\rho^{(a)}(2)\ln\rho^{(a)}(2)\right\}+\mbox{Tr}\left\{\rho\ln\rho\right\},
\end{eqnarray}
where the conditional quantum entropy for the single-qudit state is
an analog of the entropy given by Eq.~(\ref{28}) for the three-qudit
state. The difference of this entropy from zero characterizes the
difference of the density matrix of the single-qudit state from the
product of the matrices
$\rho^{(a)}(1)\otimes\rho^{(a)}(2)\otimes\rho^{(a)}(3)$. The value
$I_{Cq}^{(a)}$ is a measure of hidden quantum correlations in
single-qudit states; if it is equal to zero, the density matrix of
the single-qudit state can be presented in the product form
$\rho^{(a)}(1)\otimes\rho^{(a)}(2)\otimes\rho^{(a)}(3)=\rho$.

The notion of entanglement in the states of a single qudit can be
defined introducing an analog of the notion of entanglement in
bipartite system. We consider the state of a single qudit with the
density matrix $\rho_{yy'}$, where $y,y'=1,2,\ldots,N$, as the
separable state if it can be expressed in the form
\begin{equation}\label{51}
\rho_{y(x_1,\,x_2),\,y'(x'_1,\,x'_2)}=\sum_kp_k\left(\rho^{(a)(k)}(1)\otimes
\rho^{(a)(k)}(2)\right)_{x_1x_2,\,x'_1x'_2},
\end{equation}
with $p_k$ being any probability distribution. If the matrix
$\rho_{yy'}$ cannot be presented in such a form, we call the state
of a single qudit the entangled state. The entanglement phenomenon
in the single qudit state reflects the presence of hidden quantum
correlations. The mathematical aspects of the entanglement in the
single qudit state are identical to the mathematical aspects of the
bipartite system states of two qudits where quantum correlations are
correlations of degrees of freedom of two subsystems. The definition
of artificial multiqudit entanglement in a single-qudit state is the
straightforward generalization of the above bipartite-state
entanglement definition.

\section{Example of the Four-Level Atom}
\pst
In this section, within the framework of the approach with functions
detecting hidden correlations, we consider the example of a
single-qudit state realized by the four-level atom or qudit with
spin $j=3/2.$ The density matrix of this qudit $\rho_{mm'}=\langle
3/2,m\mid\widehat\rho\mid 3/2,m'\rangle$, where
$m,m'=-3/2,-1/2,1/2,3/2$, can be denoted as $\rho_{yy'}$
$(y,y'=1,2,3,4)$ using the map $-3/2\leftrightarrow 1$,
$-1/2\leftrightarrow 2$, $1/2\leftrightarrow 3$, $3/2\leftrightarrow
4$. The numbers used to define the functions detecting hidden
correlations in the system of two artificial qubits are as follows:
$N=4$, $X_1=2$, and $X_2=2$. Then we have for (\ref{A}) the
functions $y(x_1,x_2)$, $x_1(y)$, and $x_2(y)$, namely,
\begin{eqnarray}\label{A1}
y(x_1,x_2)=x_1+(x_2-1)X_1,\qquad x_1=1,2,\qquad x_2=1,2,\qquad
X_1=2,\nonumber\\[-2mm]
\\[-2mm]
x_1(y)=y\,\mbox{mod}\,2,\qquad
x_2(y)=\frac{y-x_1(y)}{2}\,\mbox{mod}\,2~+1,
\qquad y=1,2,3,4.\nonumber
\end{eqnarray}
These functions are described explicitly by numbers
\begin{eqnarray*}
&&y(1,1)=1,\quad y(2,1)=2,\quad y(1,2)=3,\quad y(2,2)=4,\quad
x_1(1)=1,\quad x_1(2)=2,\\
&&x_1(3)=1,\quad x_1(4)=2,\quad x_2(1)=1,\quad x_2(2)=1,\quad
x_2(3)=2,\quad x_2(4)=2.
\end{eqnarray*}
Thus, the density matrix of the qudit with $j=3/2$ can be written as
the density matrix of two two-level atoms (two artificial qubits),
i.e.,
\begin{equation}\label{31}
\rho_{yy'}\equiv\rho_{y(x_1,\,x_2),\,y'(x'_1,\,x'_2)}\equiv\rho_{x_1x_2,\,x'_1x'_2}.
\end{equation}
For example, the state of the four-level atom with the density
matrix $\rho_{yy'}=\dfrac{1}{2}\left(\begin{array}{cccc}
1&0&0&1\\
0&0&0&0\\0&0&0&0\\1&0&0&1\end{array}\right)$ can be considered as
the state with the density matrix
$\rho_{x_1,x_2,x'_1,x'_2}=\left(\begin{array}{cccc}
1/2&0&0&1/2\\
0&0&0&0\\0&0&0&0\\1/2&0&0&1/2\end{array}\right), $ 
for which the density matrices of two artificial qubits
$\rho_{x_1,x'_1}$ and $\rho_{x_2,x'_2}$ are
$\rho_{x_1,x'_1}=\rho_{x_2,x'_2}=\left(\begin{array}{cc}
1/2&0\\0&1/2\end{array}\right).$

Due to the $ppt$-criterion~\cite{Peres,Gorod}, the above state of
two artificial qubits is known to be an entangled state. Thus, the
pure state of the four-level atom has the hidden correlations
associated with the behavior of artificial qubit degrees of freedom.
Hidden correlations also exist for single qudits with $N$ given by
prime numbers. A tool to understand this property is to introduce
$\widetilde N=N+k$, where $\widetilde N=N_1N_2$, and to extend the
density matrix $\rho_{yy'}$ adding an appropriate number of zero
columns and rows. Thus, for the new density matrix $\rho_{\widetilde
y{\widetilde y}'}$ with $\widetilde y,{\widetilde
y}'=1,2,\ldots,N+k$, we repeat our construction by introducing
artificial qudits in a new (extended) Hilbert space. Thus, the
five-level atom can be considered as a bipartite system of an
artificial qubit and an artificial qutrit.

\section{Quantum Thermodynamics and Hidden Correlations}
\pst
Recently, some problems of quantum thermodynamics were discussed in
\cite{Oppenheim,Huber}; the problems are connected with different
relationships between the density matrices of thermal-equilibrium
states $\rho(\hat H,T)=\exp\,(-\hat
H/T)\,{\big/}\,\mbox{Tr}\,\exp\,(-\hat H/T)$, where $\hat H$ is the
system Hamiltonian and $T$ is a parameter, which can be considered
as the temperature, and the other density matrices (see, for
example,~\cite{Castanos}). The density matrix of the
thermal-equilibrium state contains the partition function $Z(\hat
H,T)=\mbox{Tr}\,\exp(-\hat H/T)$; this function determines the von
Neumann entropy $S(\hat H,T)=-\mbox{Tr}\,\big(\rho(\hat
H,T)\ln\rho(\hat H,T)\big)$ and free energy $F(\hat H,T)=E(\hat
H,T)-TS(\hat H,T)$, where energy $E(\hat H,T)=\mbox{Tr}\,\big(\hat
H\rho(\hat H,T)\big)$ of the thermal-equilibrium state.

In \cite{Castanos}, we obtain the new inequality for dimensionless
energy and entropy associated with any other state with a
finite-dimensional density matrix $\rho$; it reads
\begin{equation}\label{51}
E(\rho,\hat H)+S(\rho)\leq \ln Z(\hat H,T=-1),
\end{equation}
where $E(\rho,\hat H)=\mbox{Tr}\,(\hat H\rho)$,
$S(\rho)=-\mbox{Tr}\,(\rho\ln\rho)$, and the value of $T$ in the
partition function is equal to $-1$. The equality in this relation
takes place only if the density matrix $\rho$ coincides with the
density matrix of the thermal-equilibrium state~\cite{22}; the other
new inequalities were obtained in \cite{Julio}.

Our goal in this section is to discuss the notion of hidden
correlations in quantum thermodynamics.

For a single-qudit state with the density matrix
$\rho(\beta)=\exp\,(-\beta/H)\,{\big/}\,\mbox{Tr}\,\exp\,(-\beta
H)$, where $\beta$ is a parameter and $H$ is any Hermitian matrix
(for example, a Hamiltonian matrix), one can introduce the notion of
artificial qudits and corresponding hidden correlations.

In the case of a four-level atom, the matrix $\rho(\beta)$ has the
matrix elements $\rho(\beta)_{yy'}$, which can be presented in the
form
\begin{equation}\label{52}
\rho(\beta)_{yy'}=\left(\begin{array}{cccc}
\rho_{11}(\beta)&\rho_{12}(\beta)&\rho_{13}(\beta)&\rho_{14}(\beta)\\
\rho_{21}(\beta)&\rho_{22}(\beta)&\rho_{23}(\beta)&\rho_{24}(\beta)\\
\rho_{31}(\beta)&\rho_{32}(\beta)&\rho_{33}(\beta)&\rho_{34}(\beta)\\
\rho_{41}(\beta)&\rho_{42}(\beta)&\rho_{43}(\beta)&\rho_{44}(\beta)\end{array}\right),
\end{equation}
and the normalization condition $\sum_{y=1}^4\rho(\beta)_{yy}=1$
holds.

The same matrix~(\ref{52}) can be written as the density matrix of
two artificial qubits employing Eq.~(\ref{31}); this provides the
possibility to construct the density matrices of two artificial
qubits,\\[2mm]
\noindent
$\rho(\beta)_{x_1x'_1}^{(1)}=\left(\begin{array}{cc}
\rho_{11}(\beta)+\rho_{22}(\beta)&\rho_{13}(\beta)+\rho_{24}(\beta)\\
\rho_{31}(\beta)+\rho_{42}(\beta)&\rho_{33}(\beta)+\rho_{44}(\beta)\end{array}\right),
\quad \rho(\beta)_{x_2x'_2}^{(2)}=\left(\begin{array}{cc}
\rho_{11}(\beta)+\rho_{33}(\beta)&\rho_{12}(\beta)+\rho_{34}(\beta)\\
\rho_{21}(\beta)+\rho_{43}(\beta)&\rho_{22}(\beta)+\rho_{44}(\beta)\end{array}\right)$.\\[2mm]
The mutual quantum information for the four-level atom state is
introduced for a thermal-equilibrium-like state
$\rho(\beta)=\exp\,(-\beta H)\,{\big/}\,\mbox{Tr}\,\exp\,(-\beta H)$
as follows:
$$ I_q(\beta, H)=-\mbox{Tr}\,\rho^{(1)}(\beta)\ln\rho^{(1)}(\beta)
-\mbox{Tr}\,\rho^{(2)}(\beta)\ln\rho^{(2)}(\beta)+
\mbox{Tr}\,\rho(\beta)\ln\rho(\beta).$$
If the Hermitian matrix $H$ coincides with the Hamiltonian $\hat H$
and the parameter $T=\beta^{-1}$ coincides with the temperature, one
has the mutual quantum information $I_q(\beta=T^{-1},\hat H)$
characterizing the hidden correlations of two artificial qubits in
the thermal-equilibrium state of the four-level atom.

The discussed characteristics can be studied in experiments with
superconducting qudits realized in devices based on Josephson
junctions~\cite{Kiktenko1,Kiktenko2,Glushkov}. The problem of using
hidden correlations in single qudits in quantum technologies
mentioned in \cite{Light,Mexico} needs extra study.

\section{Concluding Remarks}
\pst
To conclude, we point out the main results of this discussion.

We suggested a systematic approach to introducing the notion of
hidden correlations and their cha\-racteristics in indivisible
systems (systems without subsystems). Thus, for single-qudit states,
we applied the concrete map of the density-matrix indices described
by specific functions detecting the hidden correlations. These
functions provide the partitions of the sets of natural numbers. In
view of these partitions, we present the density matrices of single
qudit states in the form of density matrices of multiqudit states.
Due to these forms of the density matrices, we introduced the notion
of entanglement for the single-qudit state. Also we obtained the
formulas for mutual information and for conditional information for
single-qudit states in terms of the functions detecting the hidden
correlations in single-qudit states.

We considered an example of the four-level atom and described
artificial qubits and their density matrices determined by the
density matrix of this indivisible system. In addition, we discussed
the thermal-equilibrium-like states of the four-level atoms and
properties of artificial qubit states related to these states.

The prospectives of using the hidden correlations in indivisible
systems in quantum technologies will be studied in future
publications.

\section*{Acknowledgments}  \pst
The comments presented in this paper are due to the discussions
during the International Workshop on Foundations of Quantum
Mechanics and Applications (Institute of Mathematical Sciences,
Universidad Aut\'onoma de Madrid, Spain,
30~January\,--\,10~February, 2017). The author thanks the Organizers
of the Workshop and especially Prof.~Alberto Ibort for invitation
and kind hospitality.

\end{document}